\documentclass[oribibl,11pt]{llncs}

\usepackage[latin1]{inputenc}
\usepackage[T1]{fontenc}
\usepackage{graphicx}
\usepackage{amssymb}
\spnewtheorem{convention}{Convention}{\itshape}{\rmfamily}
\spnewtheorem{notation}{Notation}{\itshape}{\rmfamily}

\usepackage{vmargin}
\setpapersize{A4}
\setmarginsrb{1.0in}{1.0in}{1.0in}{0.9in}{0pt}{0pt}{0pt}{0.3in}
\newcommand{\ram}{\mbox{\rm RAM}}
\newcommand{\nram}{\mbox{\rm NRAM}}

\newcommand{\ntispace}{\mbox{\rm NTISP}}
\newcommand{\dtisp}{\mbox{\rm DTISP}}
\newcommand{\contisp}{\mbox{\rm co-NTISP}}

\newcommand{\ntime}{\mbox{\rm NTIME}}
\newcommand{\dtime}{\mbox{\rm DTIME}}
\newcommand{\np}{\mbox{\rm NP}}

\newcommand{\nlin}{\mbox{\rm NLIN}}
\newcommand{\dlin}{\mbox{\rm DLIN}}
\newcommand{\vertexnlin}{\mbox{\rm Vertex-NLIN}}
\newcommand{\linplanlocal}{\mbox{\rm LIN-PLAN-LOCAL}}

\newcommand{\eso}{\mbox{\rm ESO}}

\newcommand{\structnf}{$\left\langle [n],f \right\rangle$}
\newcommand{\structcnf}{$\left\langle [cn],F \right\rangle$}

\newcommand{\risa}{\mbox{\sc Risa}}
\newcommand{\sat}{\mbox{\sc Sat}}

\newcommand{\lc}{\mbox{\sc Layered-Constraints}}
\newcommand{\hamilton}{\mbox{\sc Hamilton}}
\newcommand{\cubicsubgraph}{\mbox{\sc Cubic-Subgraph}}
\newcommand{\partition}{\mbox{\sc Partition}}
\newcommand{\knapsack}{\mbox{\sc Knapsack}}
\newcommand{\clique}{\mbox{\sc Clique}}
\newcommand{\maxindset}{\mbox{\sc Maximum Independent Set}}
\newcommand{\plancol}{\mbox{\sc Plan-$3$-Col}}
\newcommand{\plansat}{\mbox{\sc Plan-Sat}}
\newcommand{\planhamilton}{\mbox{\sc Plan-Hamilton}}

\newcommand{\rr}{${\cal R}$}
\newcommand{\pp}{${\cal P}$}

\newcommand{\ii}{${\cal I}$}

\DeclareSymbolFont{AMSb}{U}{msb}{m}{n}
\DeclareMathSymbol{\N}{\mathbin}{AMSb}{"4E}
\DeclareMathSymbol{\Z}{\mathbin}{AMSb}{"5A}
\DeclareMathSymbol{\R}{\mathbin}{AMSb}{"52}
\DeclareMathSymbol{\Q}{\mathbin}{AMSb}{"51}
\DeclareMathSymbol{\I}{\mathbin}{AMSb}{"49}
\DeclareMathSymbol{\C}{\mathbin}{AMSb}{"43}
\DeclareMathSymbol{\BL}{\mathbin}{AMSb}{"4C}

\title{Lower Bounds and Complete Problems in Nondeterministic Linear Time and Sublinear Space Complexity Classes}

\author{
Philippe Chapdelaine \and Etienne Grandjean}
\institute{GREYC, Université de CAEN, Bd Mar\'echal Juin, 14032 CAEN Cedex, FRANCE\\
\email{\{philippe.chapdelaine,etienne.grandjean\}@info.unicaen.fr}}

\begin{document}

\maketitle
\pagestyle{plain}
\thispagestyle{plain}

\begin{abstract}

Proving lower bounds remains the most difficult of tasks in computational complexity theory.
In this paper, we show that whereas most natural NP-complete problems belong to \nlin\
(linear time on nondeterministic \ram s), some of them, typically the planar versions
of many \np-complete problems, belong to \ntispace$(n,n^q)$, for some $q<1$, i.e.,
are recognized by nondeterministic \ram s in linear time and sublinear space.
The main results of this paper are the following: as the second author did for \nlin, we
\begin{itemize}
  \item give  exact logical characterizations of nondeterministic polynomial time-space complexity classes;
  \item derive from them a problem denoted $\lc (t,s)$, which is complete in the class
  \ntispace$(n,n^{s/t})$, for all integers $t,s$, $t \geq s \geq 1$, and
  \item as a consequence of such a precise result and of some recent separation
  theorems by \cite{fortnowmelkebeek00} using diagonalization, prove time-space
  lower bounds for this problem, e.g. $\lc(3,2) \notin \dtisp(n^{1.618},n^{o(1)})$.\\
\end{itemize}


\end{abstract}

\section{Introduction and discussion} \label{intro}

\subsection{The difficulty to prove complexity lower bounds}

One of the main goals of computational complexity is to prove lower bounds
for natural problems. In his Turing Award Lecture \cite{cook83}
twenty years ago, S. Cook noted: ``There is no nonlinear time lower bound
known on a general purpose computation model for any problem in \np,
in particular, for any of the $300$ problems listed in \cite{gareyjohnson79}''.
Since $1983$, despite of some technical progress (see for example
\cite{beame91,fortnow00,fortnowmelkebeek00,grandjean90risa,gurevich-shelah90,kannan84,paulPST83})
things have not fundamentally changed and Fortnow \cite{fortnowmelkebeek00}
wrote in $2000$: ``Proving lower bounds remains the most difficult of tasks
in computational complexity theory. While we expect problems like satisfiability
to take time $2^{\Omega(n)}$, we do not know how to prove that non linear-time
algorithms exist on random-access Turing machines''. In our opinion, the persistent
difficulty to prove time lower bounds for natural \np-complete problems is due to the
conjunction of two facts.
\begin{enumerate}
  \item[$(i)$] Nondeterminism makes such problems easy, typically they belong to \nlin,
  i.e. are recognized in linear time on nondeterministic \ram s, and most of them are
  even easier, i.e. we conjecture that they are not \nlin-complete.
  \item[$(ii)$] Almost nothing is known about the relationships between deterministic
  time and nondeterministic time.
\end{enumerate}
Let us now develop both arguments $(i)$ and $(ii)$.

\subsection{Nondeterminism makes problems easy}

In a series of papers \cite{grandjean94linear,grandjean96,grandjean-olive2003},
the second author showed that many \np-complete problems, including the $21$
problems studied in the seminal paper of Karp \cite{karp72}, belong to \nlin,
i.e. can be recognized in linear time on nondeterministic \ram s,
and \cite{grandjean90risa,grandjean94linear,ranaivoson91} (see also
\cite{grandjean-schwentick2002}) proved that a few of them are \nlin-complete,
including the problem \risa\ (Reduction of Incompletely Specified Automata:
quoted [AL$7$] in the well-known book \cite{gareyjohnson79} of Garey and Johnson).
Moreover, \cite{grandjean96} and \cite{barbanchon-grandjean2002} argue that
it is unlikely that many \np-complete problems in \nlin\ such as \sat
(the satisfiability problem) and \hamilton\ (the Hamilton cycle problem) are \nlin-complete.
Further, several authors
\cite{lipton-tarjan1, lipton-tarjan2, robson85, stearns-hunt90}
give convincing arguments that a number of \np-complete problems,
including \clique, \partition\ and planar restrictions of \np-complete problems
are even easier. E.g., \cite{lipton-tarjan2} deduced from the
Planar Separator Theorem (see \cite{lipton-tarjan1}) that the
\maxindset\ problem in planar graphs can be computed in deterministic subexponential
time $2^{O(n^{1/2})}$, whereas we believe the same result does not hold
for many other \np-complete problems including \sat\ (see \cite{stearns-hunt90}).
Finally, in the same direction, \cite{grandjean-olive2003} recently proved
that a couple of graph problems, including \hamilton\ and \cubicsubgraph\
belong to the class \vertexnlin, i.e., are recognized by
nondeterministic \ram s in time $O(n)$, where $n$ is the number of vertices
of the input graph, which may be much less than the size (number of edges)
of the graph.

\subsection{Our ignorance of the relationships between deterministic time and nondeterministic time}

Whereas most people expect \np-complete problems to be exponential,
there are only very modest results that formally prove that nondeterminism
gives strictly more power to computation. Interestingly, \cite{paulPST83}
proved that nondeterministic Turing machines (TM) compute strictly more
problems in linear time than deterministic ones, namely
$\dtime_{\rm TM}(n) \subsetneq \ntime_{\rm TM}(n)$. Using this result
and the inclusion $\ntime_{\rm TM}(n) \subseteq \nlin$, \cite{grandjean90risa}
concludes that \risa\ (or any other similar \nlin-complete problem via
$\dtime_{\rm TM}(n)$ reductions) does not belong to $\dtime_{\rm TM}(n)$.
However, it would be much more significant to obtain the similar but
stronger result for deterministic \ram s, namely $\risa \notin \dlin$,
that is equivalent to the conjecture $\dlin \ne \nlin$.
This would be a very strong result since, as argued in
\cite{schwentick97, grandjean-schwentick2002}, the class \dlin\ exactly
formalizes the important and very large class of linear time computable problems.

Despite of our pessimistic arguments ($i$-$ii$), some progress has been recently
made by considering mixed time-space complexity.

\subsection{Time-space lower bounds}

In recent years, Fortnow and several authors
\cite{fortnow00,fortnowmelkebeek00,liptonviglas99,tourlakis00} have used a
new approach to show that some problems like \sat\ require a nonminimal amount
of time or space. Their techniques inspired by some earlier work of
\cite{kannan84} essentially use two arguments sketched below.
\begin{enumerate}
  \item {\bf A hardness result:} \sat\ is ``complete'' for quasi-linear time
  $O(n(\log n)^{O(1)})$ under reductions that use quasi-linear time and logarithmic
  space (see \cite{cook88,dewdney82,schnorr78});
  \item {\bf A separation result proved by diagonalization:} There exist
  constants $a,b$ such that $\ntime(n) \not \subseteq \dtisp(n^a,n^b)$
  (see \cite{fortnowmelkebeek00}), where $\dtisp(T(n),S(n))$ denotes the
  class of problems computable on deterministic \ram s in time $O(T(n))$
  {\em and} space $O(S(n))$.
\end{enumerate}

From ($1$-$2$), Fortnow et al. \cite{fortnowmelkebeek00} conclude:
for any constants $a'<a, b'<b : \sat \notin \dtisp(n^{a'},n^{b'})$.

Finally, note that another completely different current of research
(see e.g. \cite{borodincook82,  beame91, gurevich-shelah90})uses
combinatorial techniques to prove lower bounds for specific problems.
However, to our knowledge such techniques have never been compared to
the ``hardness-separation'' method ($1$-$2$) above.

Let us now describe the contribution of this paper.

\subsection{Our contribution}

In this paper we generalize for mixed time-space complexity classes $\ntispace(n,n^{s/t})$ and
$\ntispace^\sigma(n^t,n^s)$ (for any signature $\sigma$ and any integers $t \geq s \geq 1$) the results of
\cite{grandjeanolive98, grandjean-olive2003} about \nlin\ and similar time complexity classes $\ntime^\sigma(n^t)$
for \ram s \footnote{$\ntispace^\sigma(T(n),S(n))$ denotes the class of $\sigma$-problems
(i.e. sets of first-order structures of signature $\sigma$)
recognizable by nondeterministic \ram s in time $O(T(n))$ and space $O(S(n))$ where $n$
is the cardinality of the domain of the input $\sigma$-structure. This generalizes
the notation of \cite{grandjean-olive2003}.}. The organization and the main results of the paper are the following:
in Section \ref{problems in classes} we show that many significant \np-complete problems
belong to some ``sublinear'' classes $\ntispace(n,n^q)$, $q<1$. In Section
\ref{characterization} we introduce the logic $\eso^\sigma(r,s)$ and prove the exact
characterization $\eso^\sigma(r,s) = \ntispace^\sigma(n^{r+s},n^s)$. In Section
\ref{completeness}, we obtain a problem, denoted $\lc(t,s)$, that is complete in the class
$\ntime(n,n^{s/t})$ and we deduce lower bounds for this problem in Section \ref{lower bounds}.

\section{Time-linear and space-sublinear classes contain significant problems} \label{problems in classes}

One important condition for a complexity class to be pertinent is to contain
natural problems. In this section, we show  that the class NTISP$(n,\sqrt n)$,
that trivially contains the \clique\ problem, contains many planar graph
problems and some problems over numbers. Moreover, we show that there also
are significant problems in the classes $\ntispace(n,n^{1-\frac{1}{d}})$,
for each integer $d \geq 2$.

\subsection{The case $\ntispace(n,\sqrt n)$}

The examples given in the case $\ntispace(n,\sqrt n)$ mainly concern planar graph problems.
We first need a separator result for planar graphs that allows a more convenient presentation of the input.

\begin{lemma} \cite{lipton-tarjan1} \label{sep planar graph}
Let $G$ be any $n$-vertex planar graph. The vertices of $G$ can be partitioned into
three sets $A$, $B$, $C$ such that no edge joins a vertex in $A$ to a vertex in $B$,
neither $A$ nor $B$ contains more than $2n/3$ vertices, nor $C$ contains more than
$2 \sqrt 2 \sqrt n$ vertices. Furthermore, $A$, $B$ and $C$ can be computed in time $O(n)$.
\end{lemma}

Now we can use this lemma recursively, i.e. find a separator set $S_1$ for $A$
and $S_2$ for $B$ and so on, until we have subsets of size $O(\sqrt n)$.
In such a way we build a binary tree that represents our graph so that the nodes
of the tree are the subsets $S,S_1,S_2,\ldots$ and every edge of the graph joins
two vertices in the same node or in two nodes of the same branch
(because any (separator) node disconnects its two child subgraphs).
We call this tree a {\em separating tree}.

\begin{lemma} \label{computing a separating tree}
Computing a separating tree $T$ of a planar graph $G=(V,E)$ can be done
in time $O(n \log n)$ and space $O(n)$, with $|V|=n$.
\end{lemma}

\begin{proof}
Consider the following facts.
\begin{itemize}

  \item The tree is linear in size (i.e., $O(n)$), as its nodes form a partition of the vertices of the graph.

  \item Each subset of type $A$ or $B$ (i.e., every successive subtree) is of size at
  most $(2/3)n$, $(2/3)^2n$, $(2/3)^3n$, etc. Thus, to reach size $O(\sqrt n)$, we need $O(\log n)$ steps, so the depth of the tree is $O(\log n)$.

  \item Now by Lemma \ref{sep planar graph} a separating step (computing $S$, $A$ and $B$) is done in time and space linear in the size of the subgraph involved. Considering that, at any given level of the tree, the union of the subgraphs at the nodes of that level is a subpartition of the whole graph, and so is of size $O(n)$, the whole computation of the separation of all the subgraphs at that level is done in time and space $O(n)$.

   \item Hence, the whole computation of the tree is done in time $O(n \log n)$ and space $O(n)$.

\end{itemize}
\qed
\end{proof}

For each planar graph problem, we can give a new version with this representation,
for example for the {\sc $3$-Colourability} problem.

\begin{description}
  \item[{\bf Problem} {\sc Separating Tree Planar $3$-Colourability}]
  \begin{description}
    \item
    \item[{\bf Instance :}] An undirected planar graph $G=(V,E)$ given in
    the separating tree representation $T$.

    \item[{\bf Question :}] Is $G$ $3$-colourable?
  \end{description}
\end{description}

\begin{proposition} \label{ntispn-sqrtn}
{\sc Separating Tree Planar $3$-Colourability} and
{\sc Separating Tree Planar Vertex Cover} are in NTISP$(n,\sqrt n)$.
\end{proposition}

In order to prove Proposition \ref{ntispn-sqrtn}, we need the
following additional result:

\begin{lemma} \label{lg branche}
In each separating tree of a planar graph $G=(V,E)$, each  branch
composed of nodes $E_0,E_1,\ldots,E_l$ has $|E_0| + |E_1| + \ldots
+ |E_l|=O(\sqrt n)$ vertices, where $|V|=n$.
\end{lemma}

\begin{proof}
Along a given branch, the subtrees at the successive nodes  are of
size at most $n$, $2n/3$, $(2/3)^2 n$, etc. So the size of each
separator  is successively at most $k \sqrt n$, $k \sqrt {2n/3}$,
$k \sqrt {(2/3)^2 n}$, etc. where $k=2 \sqrt 2$.  Therefore, the
size of a branch without its leaf is:
\begin{eqnarray}
|E_0|+|E_1|+\ldots+|E_{l-1}| & \leq & \sum_{i=0}^l k \sqrt{(2/3)^i n}   \nonumber \\
                         & \leq & k \sqrt n \sum_{i=0}^\infty {\sqrt{(2/3)}}^i  \nonumber \\
                         & = & O(\sqrt n) \nonumber
\end{eqnarray}

Finally, the subset $E_l$ at the leaf is, by definition of  the
separating tree, of size $O(\sqrt n)$ so overall the size
$|E_0|+\ldots+|E_l|$ of the branch is $O(\sqrt n)$.\qed
\end{proof}

\begin{proof}[of Proposition \ref{ntispn-sqrtn}]
Consider the first branch of the separating tree. It is  of size
$O(\sqrt n)$ by Lemma \ref{lg branche} and so we can guess for it
a $3$-colour assignment and check that it is correct in time and
space each $O(\sqrt n)$. Now, following a depth-first search
algorithm pattern, we can successively forget the assignment at
the leaf (we have already checked that it is correct, and no
further edge will ever lead to a vertex in this subgraph) and
process the next branch. So we can recursively check the
$3$-colourability of the graph, one branch at a time, while
visiting every node at most as often as there are edges leading to
it. Therefore at any time, there is never more than one branch in
memory, limiting the size to $O(\sqrt n)$, and the total number of
visits to nodes is at most the number of edges, which for planar
graphs is $O(n)$.

The proof is similar for {\sc Separating Tree Planar Vertex
Cover}. \qed
\end{proof}

Although there is an $O(n \log n)$ delay to build the tree,
which prevents to prove that $\plancol \in \ntispace(n,\sqrt n)$,
this proposition still has a significant consequence concerning an
upper bound for this problem.

\begin{proposition} \label{inclusion-classe}
If $\ntispace(n,\sqrt n) \subseteq \dtisp(T(n),S(n))$,
with $T(n) \geq n \log n$ and $S(n) \geq n$, then $\plancol \in \dtisp(T(O(n)),S(O(n)))$.
\end{proposition}

Moreover, a similar result can be applied to a wide range of problems,
namely those linearly equivalent to \plancol, as stated by the following results.

\begin{lemma} \cite{barbanchon-grandjean2002}
\plancol\ is linearly equivalent to \plansat\ (i.e., there are \dlin\
\cite{barbanchon-grandjean2002,grandjean-schwentick2002} reductions both
from \plancol\ to \plansat\ and from \plansat\ to \plancol) and \planhamilton.
\end{lemma}

\begin{definition} \cite{barbanchon-grandjean2002}
\linplanlocal\ is the class of problems linearly reducible to \plansat.
\end{definition}

\begin{corollary} \label{corollaire}
If $\ntispace(n,\sqrt n) \subseteq  \dtisp(T(n),S(n))$,
with $T(n) \geq n \log n$ and $S(n) \geq n$, then $\linplanlocal \subseteq \dtisp(T(O(n)),S(O(n)))$.
\end{corollary}

This last corollary shows that a result similar to Proposition \ref{inclusion-classe}
can be applied to the wide range of problems that are linearly reducible to \plansat.
Note that Corollary \ref{corollaire} is much more precise than the previously known
inclusion $\linplanlocal \subseteq \dtime(2^{O(\sqrt n)})$ \cite{barbanchon-grandjean2002}
(because of the inclusion $\ntispace(n,\sqrt n) \subseteq \dtime(2^{O(\sqrt n)})$).\\

Other interesting problems that happen to be in $\ntispace(n,\sqrt
n)$ are the well-known \partition\  \cite[ref
SP12]{gareyjohnson79} and \knapsack\ \cite[ref
MP9]{gareyjohnson79} problems.

\begin{proposition} \label{partition}
The problems \partition\ and \knapsack\ are in $\ntispace(n,\sqrt n)$.
\end{proposition}

\begin{proof}
Recall that the \partition\ and \knapsack\ problems are defined as
follows.

\begin{description}
  \item[{\bf Problem} \partition]
  \begin{description}
    \item
    \item[{\bf Instance :}] A finite set $A$ of integers.

    \item[{\bf Question :}] Is there a subset $A' \subseteq A$ such that $\sum_{a \in A'} a = \sum_{a \in A'\setminus A} a$?
  \end{description}
\end{description}

\begin{description}
  \item[{\bf Problem} \knapsack]
  \begin{description}
    \item
    \item[{\bf Instance :}] A finite set $U$, for each $u \in U$ a size $s(u) \in \N$ and a value $v(u) \in \N$, and positive integers $B$ and $K$.

    \item[{\bf Question :}] Is there a subset $U' \subseteq U$ such that $\sum_{u \in U'} s(u) \leq B$ and such that $\sum_{u \in U'} v(u) \geq~K$?

  \end{description}
\end{description}

The idea of the proof that \partition\ belongs to
$\ntispace(n,\sqrt n)$ is based on the one given by Hunt and
Stearns in \cite{stearns-hunt90} for $\dtime(2^{O(\sqrt n)})$.
Consider an instance $A= \left\{ a_1,\ldots,a_k \right\}$ of
\partition. The size of the input is $n$, that is the $a_i$ are
written in base $n$, and they occupy $n$ registers. Consider a
fixed real $d$, $0 \leq d <1$, and compute $n^d-1$. Consider two
empty sets $A_1$ and $B_1$, with $S(A_1)$ and $S(B_1)$ the sums of
the integers in $A_1$ and $B_1$ respectively. Now for each $a_i$
whose size is smaller or equal to $n^d-1$ registers,
nondeterministically put it in $A_1$ or in $B_1$ and add its value
to $S(A_1)$ or $S(B_1)$. Note that since there are at most $n$
numbers $a_i$ whose size is smaller or equal to $n^d-1$ registers,
then $S(A_1) \leq  n \cdot n^{n^d-1} = n^{n^d}$ (the same for
$S(B_1)$) and can be stored in $n^d$ registers. Now we consider
all the $a_i$s of size greater than $n^d-1$ and we
nondeterministically partition them into two subsets $A_2$ and
$B_2$. There clearly are no more than $n^{1-d}$ such numbers and
so this is the number of registers needed to keep a record of the
partition. Once this is done, we sum the units (in base $n$) of
the numbers in $A_2$ with the unit digit of $S(A_1)$, and we make
sure it is equal modulo $n$ to the sum of the units of the numbers
in $B_2$ and the unit digit of $S(B_1)$. This only takes $O(1)$
registers as we work in base $n$. We also compute the carry and
then we do the same for the tens, the hundreds, etc (in base $n$).
If at every stage, the sums are equal, we have a partition of
$\left\{ a_1,\ldots,a_k \right\}$.
Finally, the whole computation uses linear time and space $O(n^d+n^{1-d})$, that is $O(n^{1/2})$ if we set $d=1/2$.\\

The proof is similar for the \knapsack\ problem.\qed

\end{proof}

\subsection{Classes $\ntispace(n,n^{1-\frac{1}{d}})$}

The following parameterized problem shows that each class $\ntispace(n,n^{1-\frac{1}{d}})$,
for every integer $d \geq 2$, contains a (quite natural) problem.

\begin{description}
  \item[{\bf Problem} {\sc $d$-Constraint Tiling}]
  \begin{description}
    \item
    \item[{\bf Instance :}] An integer $d \geq 2$, $d$ integers $m_1,m_2,\ldots,m_d$ and
    a $d$-dimensional $m_1 \times m_2 \times \ldots \times m_d$ grid, with a set of
    allowed tiles for each hypercube of the grid. Each tile has its faces coloured.

    \item[{\bf Question :}] Can we choose for every hypercube of the grid one of its
    allowed tiles so that two adjacent hypercubes have the same colour on their common face?
  \end{description}
\end{description}

\begin{proposition} \label{d-tiling}
For every integer $d \geq 2$, the problem {\sc $d$-Constraint Tiling} is in $\ntispace(n,n^{1-\frac{1}{d}})$.
\end{proposition}

\begin{proof}
We prove this result for $d=2$, the general case being an easy
generalization. Consider a rectangle consisting of $n \times m$
squares, with each one having a nonempty set of tiles. The size of
the input is  $t \geq n m$. Suppose that $n \geq m$. Consider the
first row of $m$ squares and choose nondeterministically an
allowed tile for each square. Now do the same for the second row
and check that both row are consistent internally and with each
other. Once this is done, the choices made for the first row are
of no more use and can be forgotten. The memory space they
occupied can be used to hold the tiles chosen for the third row,
then checking the consistence with the second row. And so on,
until we have in memory the $n-1^{th}$ and $n^{th}$ rows. At any
time we only keep $2$ rows  in memory, each of size $m$, and the
space used is always the same recycled, that is $O(\sqrt t)$
(recall that  $m \leq \sqrt {mn} \leq \sqrt t$), and it is easy to
see that the whole process takes time $O(t)$. \qed
\end{proof}

\section{Computational and logical preliminaries} \label{preliminaries}

\subsection{\nram s and time-space complexity classes}

The time-space functions studied here being very tight, it is very important to
describe precisely the computational model we use, that is the Nondeterministic
Random Access Machine (or \nram) as it was designed by Grandjean and al.
in several papers (see for example \cite{grandjeanolive98, schwentick97, grandjean-olive2003}),
with only slight modifications.\\

\begin{figure}[htbp]
\begin{center}
\includegraphics[width=8cm]{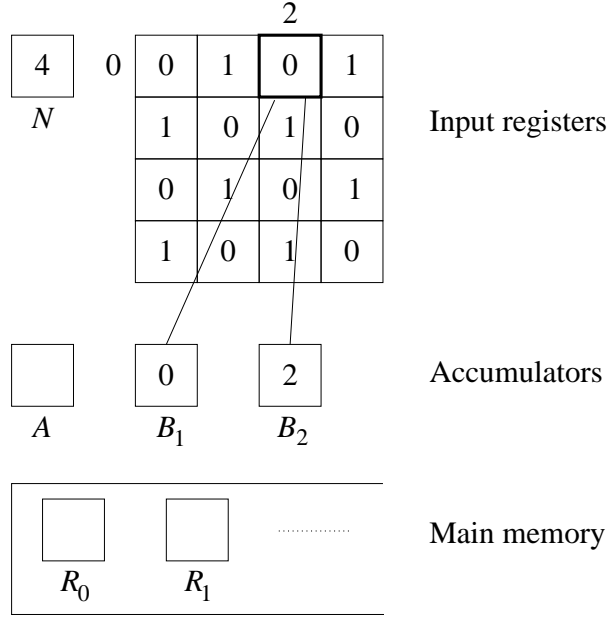}
\end{center}
\caption{An \nram\ associated with a binary relation} \label{ram}
\end{figure}

An \nram\ \rr\ is designed to store an input structure $S=\langle
[n],\sigma \rangle$, where $[n]= \{ 0,1,\ldots,n-1 \}$ and
$\sigma$ is a finite signature containing $p$ function or
predicate symbols \footnote{In our notation, we confuse each
signature (resp. function or predicate symbol) with its
interpretation.}. It consists of (see Figure \ref{ram}):

\begin{itemize}
  \item {\em input registers}: a register $L$ containing the integer $n$,
  and for each $\sigma$-symbol $f$ of arity $k$, and for each tuple $a \in [n]^k$,
  a register $f[a]$ containing the value of $f$ in $a$;
  \item {\em the working section} composed of {\em $d+1$ special registers}
  (called {\em accumulators}), $A,B_1,\ldots,B_d$, where
  $d=max_{f \in \sigma} \{ arity(f) \}$, and {\em the main memory} which
  consists of computation registers $R_0,R_1,\ldots$

\end{itemize}

\begin{convention}
\begin{itemize}
  \item The input registers are called $Q_j(a)$, where $Q_j$ is the $j^{th}$
  symbol of $\sigma$, $1 \leq j \leq p$, and $a \in [n]^k$, where $p$ is the
  number of symbols of $\sigma$ and $k$ is the arity of $Q_j$.

  \item All the input registers are read-only while the computation registers
  $A,B_1,\ldots,B_d,R_0,\dots$ are read/write.
\end{itemize}
\end{convention}

The program of the \nram\ \rr\ is a sequence of instructions
${\cal I}(1),{\cal I}(2),\dots,{\cal I}(\lambda)$ of the following types ($1 \leq j \leq p , 1 \leq i \leq d$):
$$
\begin{tabular}{rlcrl}
$(1)$ & $A:=L$                      & $\quad$ & $(8,i)$  & $B_i:=A$ \\
$(2)$ & $A:=0$                      &         & $(9,i)$  & $R(A):=B_i$ \\
$(3)$ & $A:=A+1$                    &         & $(10,i)$ & $if$ $A=B_i$ $then$ $goto$ ${\cal I}(i_0)$ \\
$(4)$ & $A:=A-1$                    &         &           & \multicolumn{1}{r}{$else$ $goto$ ${\cal I}(i_1)$}\\
$(5)$ & $guess(A)$                  &         & $(11)$    & $accept$ \\
$(6,j)$ & $A:=Q_j(B_1,\ldots,B_k)$  &         & $(12)$    & $reject$ \\
$(7)$ & $A:=R(A)$                   &         &           & \\
\end{tabular}
$$

\begin{convention}
\begin{itemize}
   \item At the beginning of the computation, all the accumulators and the registers
   in the main memory contain the value $0$.
  \item $guess(A)$ is the non-deterministic instruction of the \nram; it stores any integer in accumulator $A$.
  \item The only $accept$ instruction in the program is \ii$(\lambda)$, that is the last one.
\end{itemize}
\end{convention}

\begin{remark}
The access to the main memory is only possible via accumulator $A$.
\end{remark}

Following this definition of our computational model, we can now define the mixed
time-space complexity classes we study here:

\begin{definition}
Let $\sigma$ be a signature and $T,S: \N \rightarrow \N$ be functions such that
$S(n) \leq T(n)$ and $T(n) \geq n$. We call $\ntispace^\sigma(T(n),S(n))$ the
class of problems over $\sigma$-structures (or $\sigma$-problems) computable
on an \nram\ using time $O(T(n))$ (i.e. that performs $O(T(n))$ instructions)
and space $O(S(n))$ (i.e. the registers of the main memory used have adresses
$O(S(n))$ and their contents  are $O(\max \{n,S(n)\})$), where $[n]$ is the domain of the input $\sigma$-structure.
\end{definition}

\begin{notation}
We will write $\ntispace(T(n),S(n))$ as an abbreviation for
$\ntispace^\sigma(T(n),S(n))$ when $\sigma$ is a unary $\{f\}$-signature,
i.e. $f$ is a unary function symbol. This corresponds to the usual convention since such a structure has size $n$.
\end{notation}

\subsection{Formulas and logical classes}

We use the standard definitions of logic and finite model theory, see e.g. \cite{ebbinghaus-flum}.

Let $succ$ be the predefined non-cyclic successor over $[n]$, that is the function
$$
\left\{
\begin{tabular}{rcll}
$y$   & $\mapsto$ & $y+1$ & if $0 \leq y <n-1$ \\
$n-1$ & $\mapsto$ & $n-1$ & \\
\end{tabular}
\right.
$$
For every $\delta \geq 1$, we define the non-cyclic lexicographical successor
function over $[n]^{\delta}$ as the following abbreviation, also denoted $succ(y_1,\dots,y_\delta)$:
$$
succ^{(\delta)}(y_1,\dots,y_\delta)=\left\{
\begin{tabular}{llll}
\multicolumn{3}{l}{$(y_1,\dots,y_{i-1},y_i+1,0,\dots,0)$} & if $(y_1,\dots,y_\delta)$ is not the \\
$\ \ \ \ \ \   $  &\multicolumn{3}{l}{last $\delta$-tuple, i.e. if for some $i=1,\dots,\delta$, we} \\
                  &\multicolumn{3}{l}{have $y_j=n-1$ for each $j>i$ and {$y_i<n-1$}} \\
                  & \\
\multicolumn{2}{l}{$(n-1,\dots,n-1)$} &                   & otherwise \\
\end{tabular}
\right.
$$

Definitions \ref{def restricted formula} and \ref{def esosr} are the cornerstone
of the main result, as they show the restrictions we impose on the logic we will
use to characterize our complexity classes.

\begin{definition} \label{def restricted formula}
A first-order quantifier-free formula $\psi(x_1,\dots,x_s,y_1,\dots,y_r)$ of
signature $\sigma \cup \left\{ 0,succ \right\} \cup \bar g$ over $s+r$ variables
is called an {\em $(s,r)$-restricted formula with input $\sigma$} if all the function
(resp. relation) symbols of $\bar g$ are of arity at most $s+r$, with the following
restriction on the arguments of those of arity $s+\delta$ ($\delta \geq 1$):
the first $s$ arguments are not restricted and the last $\delta$ arguments form a
vector of form either $(y_1,\dots,y_\delta)$ or $succ^{(\delta)}(y_1,\dots,y_\delta)$.
\end{definition}

\begin{remark}
Note that if $r=1$, %this means
the last argument of an $(s+1)$-ary function is $y$ or $succ(y)$.
\end{remark}

We can now define the classes of logical formulas that will characterize our mixed time-space complexity classes.

\begin{definition} \label{def esosr}
We call $\eso^\sigma(s,r)$ the class of  Existential Second Order formulas of the form
$$
\exists \overline{g} \ \forall x_1 \dots \forall x_s \ \forall y_1 \dots \forall y_r \ \psi(\overline{x},\overline{y})
$$
where $\overline{g}$ is a set of function or relation symbols of arity at most $s+r$, $y_1,\dots,y_r$
are called the {\em iteration variables}, and $\psi$ is a quantifier-free $(s,r)$-restricted
formula with input $\sigma$ and of signature
$\sigma \cup \left\{ 0,succ \right\} \cup \left\{ \overline{g} \right\} $.
\end{definition}

\section{Logical characterization of mixed time-space classes} \label{characterization}

A main result of this paper is the following exact characterization of each time-space
complexity class $\ntispace^\sigma(n^t,n^s)$, for all integers $t \geq s \geq 1$,
which generalizes the similar characterization of the classes $\ntime^\sigma(n^t)$ \cite{grandjean-olive2003}:

\begin{theorem} \label{th main}

For all integers $s \geq 1, r \geq 0$ and any signature $\sigma$, a $\sigma$-problem \pp\
is in $\ntispace^\sigma(n^{s+r},n^s)$ iff there exists a formula $\phi$
in $\eso^\sigma(s,r)$ that characterizes \pp, i.e. such that for every
$\sigma$-structure $\langle [n],\sigma \rangle$ of domain $[n]$:
\begin{eqnarray} \label{mainresult}
\left\langle [n],\sigma  \right\rangle \in {\cal P} \mbox{ iff }
 \left\langle [n] ,\sigma,succ,0 \right\rangle \models \phi
\end{eqnarray}

\end{theorem}

For the sake of simplicity and without loss of generality, we will
restrict ourselves to signatures $\sigma=\left\{ f \right\}$
containing only one function symbol $f$, of any arity $d$. That
means the program of the \nram\ will contain instructions $(6,j)$
of the unique following form:
$$
(6) \quad A:=f(B_1,\ldots,B_d)
$$
Also we will prove the theorem only for the case $r=s=1$ the
general case being just an easy generalization of this particular
one. Note that we will use the linear order $<$ over domain $[n]$
as it is definable in $\eso^\emptyset(1,1)$ (see
\cite{grandjean90spectra} or \cite[Corollary
2.1]{grandjean-olive2003}).
\\

First, if there exists a formula $\phi$ in $\eso^\sigma(1,1)$ such
that the equivalence above holds for every $\sigma$-structure,
then it is easy to see that \pp\ is in $\ntispace^\sigma(n^2,n)$.

Let \pp\ be a $\sigma$-problem. Suppose that there exists an
$\eso^\sigma(1,1)$  formula $\phi$ such that the equivalence $(1)$
of Theorem \ref{th main} holds for every $\sigma$-structure
$\langle [n],\sigma \rangle$. An \nram\ \rr\ can check $\phi$ in
the following way:

\begin{itemize}

\item \rr\ first $guesses$ and stores the $p_1$ unary \eso\ functions
$g_i:[n] \rightarrow [n]$, and the $2 \times (p-p_1)$ unary
restrictions of the binary \eso\ functions $g_i:[n] \times [n]
\rightarrow [n]$ by setting $y=0$ and $y=1$. All this can be done
in linear time $O(n)$ and by using a linear number of registers.
We can then check the formula for all $x$ and for $y=0$ (remember
the form of a $(1,1)$-restricted formula).

\item We now have the following loop: for $y=1$ to $y=n-1$, replace in the
registers the values of the binary functions for $y-1$ by the
values for $y$, and then $guess$ the values for $y+1$ and store
them in the registers just freed. Check whether the formula holds
for all $x$ and for the current $y$. This is done also in linear
time; the space used is the same as in the first step, so it is
still linear.

\end{itemize}

We have $n$ such iterations (including the one for $y=0$), each of
time $O(n)$, so the time used overall is quadratic. The space used
is always the same, that is linear. So we have ${\cal P} \in
\ntispace^\sigma(n^2,n)$.

Now let us see how we can describe a problem in
$\ntispace^\sigma(n^2,n)$ with a formula in $\eso^\sigma(1,1)$. A
problem \pp\ is in $\ntispace^\sigma(n^2,n)$ iff it is recognized
by an \nram\ \rr\ that works in time at most $cn^2$ and space
$cn$, and uses numbers at most $cn$, for some fixed integer $c$.
We can also suppose that if our \nram\ works in time less than
$cn^2$, then the final configuration is repeated until instant
$cn^2$ so that \rr\ works in time exactly $cn^2$ and the instants
of the computation can be labelled $0,1,2,\dots,cn^2-1$.

Our goal is to describe the computation of \rr\ with a logical
formula. As we cannot describe the content of every register at
any time (this would require a size $\Theta(\mbox{time} \times
\mbox{space})=\Theta(n^3)$, i.e. some ternary function on the
domain $[n]$), we only encode what may change: the current
instruction index, the contents of accumulators
$A,B_1,\ldots,B_d$, and the content of the register pointed to by
$A$. We want a logical formula over the domain $[cn]$, so we will
code the instants of the computation $0,1,\dots,cn^2-1$ with
ordered pairs $(t,T)$, $0 \leq t < cn$, $0 \leq T < n$, so that
$(t,T)$ encodes the instant $t + T \cdot cn$ of a computation of
\rr. Let us introduce the following functions:

\begin{itemize}
\item $I(t,T)$ denotes the index of the instruction performed at instant $(t,T)$.
\item $A(t,T)$ denotes the content of accumulator $A$ at instant $(t,T)$.
\item $B_i(t,T)$, $1 \leq i \leq d$, denotes the content of accumulator $B_i$ at instant $(t,T)$.
\item $R_A(t,T)$ denotes the content of register $R(A)$ (ie. the register pointed to by $A$) {\em at} instant $(t,T)$.
\item $R'_A(t,T)$ denotes the content of the same register $R(A)$ {\em after} step $(t,T)$.
\end{itemize}

Let us mention two things. Firstly, the encoding of the time
naturally divides the time-space diagram of the computation of
\rr\ into $n$ blocks of $cn$ instants. Secondly, all those
functions are binary and should respect the conditions of
Definitions \ref{def restricted formula} and \ref{def esosr} for
the logic $\eso^\sigma(1,1)$. This compels us to consider only two
blocks at once : the current one referred at by $T$, and the
previous one ($T-1$). So, $T$ is our {\em unique iteration
variable}. Now, remember that at any instant $t$ of block $T$, we
must be able to know the contents of accumulators $A$,
$B_1,\ldots,B_d$ and of the register pointed to by $A$. The
successive contents of the accumulators are completely described
by the above functions $A$ and $B_1,\ldots,B_d$, so there is no
problem for them. In contrast, the contents of the computation
registers are accessible only through the function $R_A$, which
must hold the right value at any time, considering the fact that
the register pointed to by $A$ at instant $(t,T)$ may be distinct
from the one at the previous instant. Moreover, if a specific
register is not accessed  in two consecutive blocks, the
restriction imposed to the iteration variable $T$ seems to prevent
the recovery of its content. So the problem is to be able to get
the value that was stored in any register the last time it was
accessed, be it in the current block or in any other block before.
The idea to overcome this difficulty is to resume, at the
beginning of each block $T$ ($0 \leq T <n$), the content of any
computation register of address $x$ ($0 \leq x <cn$) with a binary
function $R(x,T)$. More precisely, $R(x,T)$ will code the content
of register $R(x)$ at the instant $(0,T)$, that is the instant
$cn\cdot T$ of the computation of \rr. We are now ready to give
the formulas that describe the computation of \rr. First, the
initial conditions are described by formula $\phi_1$:
$$
\phi_1 \equiv I(0,0)=1 \ \wedge \ A(0,0)=0 \ \wedge \ B_1(0,0)=0 \wedge \ldots \wedge B_d(0,0)=0
$$

Functions $I$, $A$, $B_i$, $1 \leq i \leq d$, and $R'_A$ can be
easily defined by recurrence from $I$, $A$, $B_i$ and $R_A$ by
$\eso^\sigma(1,1)$ formulas $\phi_I$, $\phi_A$, $\phi_{B_i}$ and
$\phi_{R'_A}$ respectively.

We use the following conventions:

\begin{itemize}

\item We use two different successor functions. The first one is the
one described above ($succ^{(1)}$) and will be applied to $T$. The
second one is the successor function over $[cn]$ and it will be
applied to $t$. Both are denoted by $succ$ since they have roughly
the same behaviour. Moreover, we use the abbreviation $succ(t,T)$
for:
$$
succ(t,T)=\left\{
\begin{tabular}{ll}
$(succ(t),T)$ & if $t<cn-1$ \\
$(0,succ(T))$ & if $t=cn-1$ and $T<n-1$ \\
$(cn-1,n-1)$  & otherwise \\
\end{tabular}
\right.
$$

\item The input function $f:[n]^d \rightarrow [n]$ is padded by the function
$$
\begin{tabular}{rrcl}
$F:$ & $[cn]^d$           & $\rightarrow$ & $[cn]$ \\
     & $(x_1,\ldots,x_d)$ & $\mapsto$     & $\left\{
\begin{tabular}{l}
$f(x_1,\ldots,x_d)$ if $x_i<n$ for all $i \leq d$,  \\
$0$ otherwise \\
\end{tabular}
\right.$ \\
\end{tabular}
$$
This function will be the only one in the input signature of the
$(1,1)$-restricted formula, and so will not be restricted in its
arguments.

\item We use the function $pred$ (non-cyclic predecessor) easily definable in $\eso^\sigma(1,1)$.

\item $I(t,T)=(i)$ (for $i=1,\dots,12$) is an abbreviation for the
disjunction ${\bigvee}_{j \in S_i}I(t,T)=j$ where $S_i \subseteq
\left\{ 1,2,\dots,\lambda \right\}$ denotes the set of indices of
instructions of type $(i)$ in the program of \rr.

\end{itemize}

We have the following case definitions:

$I(succ(t,T))=\left\{
\begin{tabular}{ll}
$i_0$          & if $I(t,T)=(10,i)$ and $A(t,T)=B_i(t,T)$ \\
$i_1$          & if $I(t,T)=(10,i)$ and $A(t,T) \neq B_i(t,T)$ \\
$I(t,T)$       & if $I(t,T)=(11)$ or $(12)$ \\
$succ(I(t,T))$ & otherwise \\
\end{tabular}
\right.$\newline \newline

$A(succ(t,T))=\left\{
\begin{tabular}{ll}
$n$            & if $I(t,T)=(1)$ \\
$0$            & if $I(t,T)=(2)$ \\
$succ(A(t,T))$ & if $I(t,T)=(3)$ \\
$pred(A(t,T))$ & if $I(t,T)=(4)$ \\
$G(t,T)$       & if $I(t,T)=(5)$ \\
$F(B_1(t,T),\ldots,B_d(t,T))$    & if $I(t,T)=(6)$ \\
$R_A(t,T)$     & if $I(t,T)=(7)$ \\
$A(t,T)$       & otherwise \\
\end{tabular}
\right.$ \newline

\noindent where the non-deterministic feature of the instruction
$guess(A)$ is given by the \eso\ function $G(t,T)$.\newline

$B_i(succ(t,T))=\left\{
\begin{tabular}{ll}
$A(t,T)$ & if $I(t,T)=(8,i)$ \\
$B_i(t,T)$ & otherwise \\
\end{tabular}
\right.$ \newline \newline

$R'_A(succ(t,T))=\left\{
\begin{tabular}{ll}
$B_i(t,T)$   & if $I(t,T)=(9,i)$ \\
$R_A(t,T)$ & otherwise \\
\end{tabular}
\right.$
\newline \newline

These case definitions of $I$, $A$, $B_i$ and $R'_A$ can be easily
transformed into first-order formulas $\phi_I$, $\phi_A$,
$\phi_{B_i}$ and $\phi_{R'_A}$ respectively, of the form $\forall
T<n \ \forall t \ \psi(t,T)$ where $\psi$ is some
$(1,1)$-restricted formula.

 So the computation of \rr\
between two successive instants will be expressed by $\phi_2$:
$$
\phi_2 \equiv \phi_I \ \wedge \ \phi_A \ \wedge \ \phi_{B_1}
\wedge \ldots \wedge \phi_{B_d} \ \wedge \ \phi_{R'_A}
$$

\begin{figure}[htbp]
\begin{center}
\includegraphics[width=6cm]{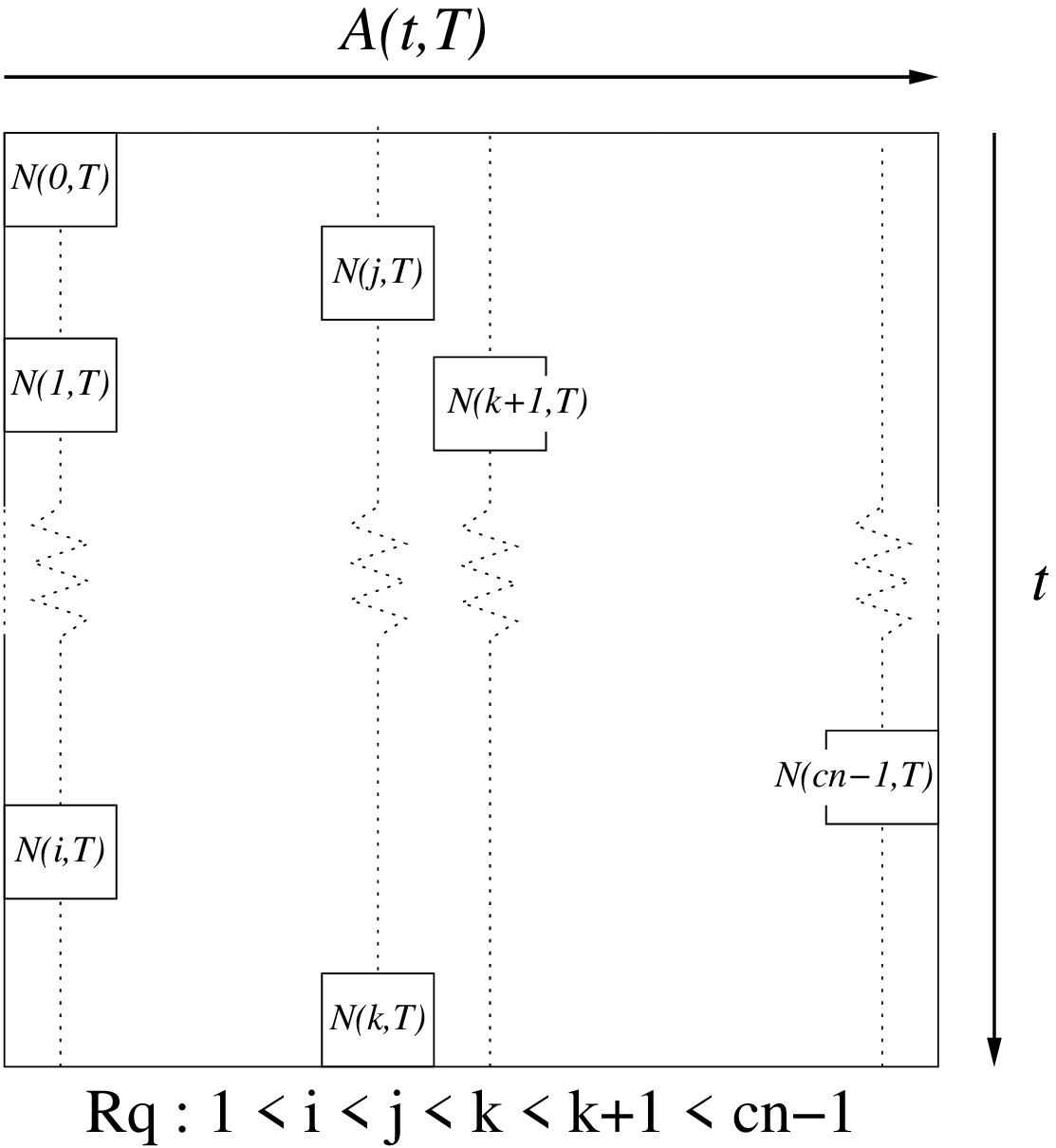}
\end{center}
\caption{$N(x,T)$} \label{nxt}
\end{figure}

Now there remains to describe functions $R_A$ and $R$, which is a
much more tricky task. For this purpose, we introduce, as in
\cite{grandjeanolive98}, the binary function
$N(x,T)=(N_1(x,T),N_0(x,T))$ (more precisely two binary \eso\
function symbols $N_0,N_1$) which represents, in each block $T$,
the lexicographical numbering of the ordered pairs $(A(t,T),t)$
(see Figure \ref{nxt} for an example on a given block $T$):

$$
\begin{tabular}{rl}
$\phi_N \ \equiv$ & $\forall T<n \ \forall t \ \exists x$ \\
                  &
\begin{tabular}{rl}
         & $N_1(x,T)=A(t,T) \ \wedge \ N_0(x,T)=t$ \\
$\wedge$ & $x \neq cn-1 \ \rightarrow \ N(x,T) <_{lex} N(succ(x),T)$ \\
\end{tabular} \\
\end{tabular}
$$
where $(i,j) <_{lex} (i',j')$ abbreviates $i<i' \vee (i=i' \wedge j<j')$.\\

Notice that the first two conjuncts of $\phi_N$ express that, for
every $T<n$, the mapping $x \mapsto N(x,T)$ is a surjection, and
hence is a bijection from the set $[cn]$ to the set of equal
cardinality $\left\{ (A(t,T),t):t \in [cn] \right\}$.

\begin{remark}
We use the non-cyclic successor functions over $[cn]$ and over
$[n]$. Both are  denoted $succ$ as they have the same behaviour.
\end{remark}

The binary  function $R$ that allows to represent the content
$R(x,T)$ of register $R(x)$ at instant $(0,T)$ is described by
formula $\phi_R$:

$$
\begin{tabular}{rl}
$\phi_R \ \equiv$ & $\forall T<n-1 \ \forall x \ \exists t \ \exists z \ \exists t' \ \exists u$ \\
                  & $\left\{ T=0 \rightarrow R(x,T)=0 \right\}$ \\
                  & $\bigwedge \ \left\{
\begin{tabular}{ll}
          & $\left\{
\begin{tabular}{ll}
            & $A(t,T)=x \ \wedge \ N(z,T)=(A(t,T),t)$ \\
$\wedge$    &
$\left[ z=cn-1 \ \vee \ \left(
\begin{tabular}{ll}
         & $z \neq cn-1$ \\
$\wedge$ & $N(succ(z),T)=(A(t',T),t') $ \\
$\wedge$ & $A(t,T) \neq A(t',T)$ \\
\end{tabular}
\right) \right]$ \\
$\wedge$    & $ R(x,succ(T))=R'_A(t,T)$ \\
\end{tabular}
\right\}$ \\
          & \\
$\bigvee$ & $\left\{
\begin{tabular}{ll}
            & $N(0,T)=(A(t,T),t)$ \\
$\wedge$ & $\left[ A(t,T)>x \ \wedge \ R(x,succ(T))=R(x,T) \right]$ \\
\end{tabular}
\right\}$ \\
          & \\
$\bigvee$ & $\left\{
\begin{tabular}{ll}
            & $N(cn-1,T)=(A(t,T),t)$ \\
$\wedge$ & $\left[ A(t,T)<x \ \wedge \ R(x,succ(T))=R(x,T) \right]$ \\
\end{tabular}
\right\}$ \\
          & \\
$\bigvee$ & $\left\{
\begin{tabular}{ll}
         & $A(t,T)<x<A(u,T) \wedge \ (A(t,T),t)=N(z,T)$ \\
$\wedge$ & $(A(u,T),u)=N(succ(z),T)$ \\
$\wedge$ & $R(x,succ(T))=R(x,T)$ \\
\end{tabular}
\right\}$ \\
\end{tabular}
\right\}$ \\
\end{tabular}
$$

\begin{remark}
The first  line of the matrix of $\phi_R$ (first conjunct)
describes the behaviour of $R$ in the first block (labelled $0$).
In the big second part (second conjunct), the first disjunct
corresponds to the case when register $R(x)$ is accessed in block
$T$ (in particular, the formula in brackets $[\dots]$ combined
with the condition $A(t,T)=x$ expresses that $(t,T)$ is the last
instant in block $T$ when $A(t,T)=x$). The other three disjuncts
correspond to the three cases when register $R(x)$ it is not
accessed in block $T$. See Figure \ref{rxsucct}  for more details.

\begin{figure}[htbp]
\begin{center}
\includegraphics[width=12cm]{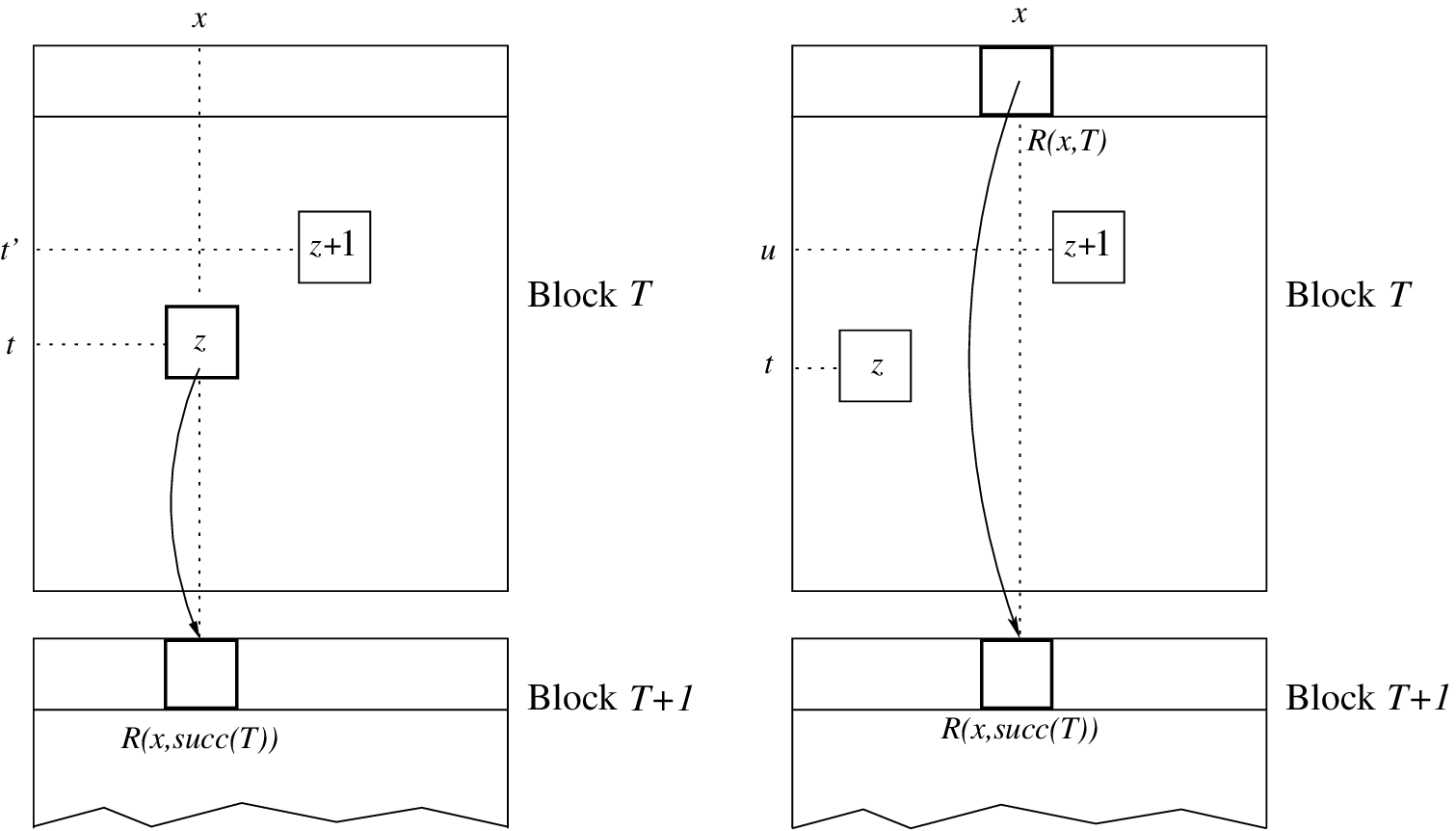}
\end{center}
\caption{$R(x,succ(T))$} \label{rxsucct}
\end{figure}
\end{remark}

By using  functions $R$ and $R'_A$, function $R_A$ (which
describes the right content of the register pointed to by
accumulator $A$) is easily defined by formula $\phi_{R_A}$ (see
Figure \ref{ratt}):

\begin{figure}[htbp]
\begin{center}
\includegraphics[width=12cm]{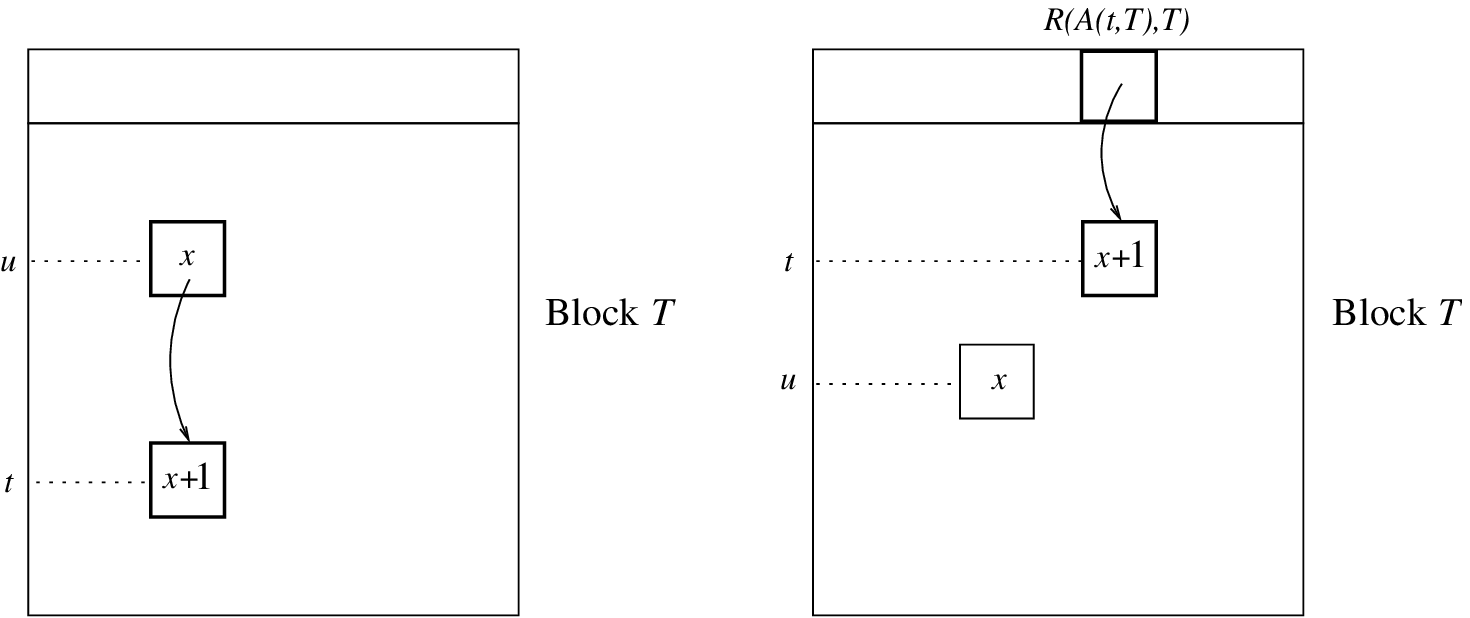}
\end{center}
\caption{$R_A(t,T)$} \label{ratt}
\end{figure}

$$
\begin{tabular}{rl}
$\phi_{R_A} \ \equiv$ & $\forall T<n \ \forall t \ \exists x \ \exists u$ \\
                      &
\begin{tabular}{ll}
          & $\left\{ (A(t,T),t)=N(0,T) \ \wedge \ R_A(t,T)=R(A(t,T),T) \right\}$ \\
$\bigvee$ & $\left\{
\begin{tabular}{ll}
         & $(A(t,T),t)=N(succ(x),T) \ \wedge \ (A(u,T),u)=N(x,T)$ \\
$\wedge$ & $\left( A(t,T)=A(u,T) \ \rightarrow \ R_A(t,T)=R'_A(u,T) \right)$\\
$\wedge$ & $\left( A(t,T) \neq A(u,T) \ \rightarrow \ R_A(t,T)=R(A(t,T),T) \right)$\\
\end{tabular}
\right\}$ \\
\end{tabular}\\
\end{tabular}
$$

So we control the contents of all the registers via formula $\phi_3$:
$$
\phi_3 \equiv \phi_N \ \wedge \ \phi_R \ \wedge \ \phi_{R_A}
$$

The fact that \rr\ reaches the {\it accept} instruction, that is \ii$(\lambda)$, is ensured by formula $\phi_4$:
$$
\phi_4 \equiv I(cn-1,n-1)=\lambda
$$

Finally, the computation of \rr\ is exactly described by formula $\phi$ over domain $[cn]$:
$$
\phi \equiv \exists I,A,B_1,\ldots,B_d,R_A,R'_A,G,N \ \phi_1 \ \wedge \ \phi_2 \ \wedge \ \phi_3 \ \wedge \ \phi_4
$$
and it is  easy to transform $\phi$ into a prenex Skolemized
$\eso^\sigma(1,1)$ formula.

We have described the computation of the \nram\ \rr\ on an input
$f$ by an \eso$(1,1)$ formula for the structure \structcnf, that
means over domain $[cn]$; now let us see how to get a formula for
the input structure \structnf, i.e. over domain $[n]$. The idea is
to code an element $x \in [cn]$ by an ordered pair of elements
$(x_0,x_1) \in [n] \times [c]$. According to this idea, every
binary function $g:[cn]\times[n] \rightarrow [cn]$ will be coded
by $2c$ functions ${g_0}^{(i)}:[n]\times[n] \rightarrow [n]$ and
${g_1}^{(i)}:[n]\times[n] \rightarrow [c]$, for $0 \leq i < c$,
defined as follows:

$$
\begin{tabular}{rcl}
${g_0}^{(i)}:[n]\times[n]$ & $\rightarrow$ & $[n]$ \\
$(x,y)$                    & $\mapsto$     & $g(in+x,y) \bmod n$ \\
                           &               & \\
${g_1}^{(i)}:[n]\times[n]$ & $\rightarrow$ & $[c]$ \\
$(x,y)$                    & $\mapsto$     & $\left\lfloor {g(in+x,y) \over n} \right\rfloor$ \\
\end{tabular}
$$

So, for every $g:[cn]\times[n] \rightarrow [cn]$, every $y \in
[n]$ and every $(x_0,x_1) \in [n] \times [c]$, we have
$g(nx_1+x_0,y)=n \cdot {g_1}^{(x_1)}(x_0,y)+{g_0}^{(x_1)}(x_0,y)$.
We notice that the iteration variable, that is $y=T$, is not
modified and hence our binary functions respect our restricted
logic (see Definitions \ref{def restricted formula} and \ref{def
esosr}). The process is similar for any unary function $g:[cn]
\rightarrow [cn]$. The details of the proof are left to the reader
\cite{grandjeanolive98}. \qed

Now, the computation of \rr\ is exactly described by formula
$\phi$ over domain $[n]$.

\section{Completeness results for some  logical problems} \label{completeness}

Before presenting our problems, along with some form of
completeness for linear time and sublinear space complexity
classes, we need some technical tools.

\subsection{A technical result}

It will  be convenient to encode any set of unary $\left\{ f
\right\}$-structures ${\cal P} \in \ntispace(m^{t/d},m^{s/d})$ -
where $s,t,d$ are fixed integers such that $t \geq  s \geq 1$ and
$t \geq d \geq 1$ and $m$ denotes the size of the unary input
structure $\langle [m],f \rangle$ - into a set ${\cal P}^{code}$
of $d$-ary structures.

\begin{remark}
We use in that notation the letter $m$ instead of $n$ to make the following encoding easier.
\end{remark}

For  fixed numbers $t,s$ with $t \geq  s \geq 1$ and any signature
$\sigma$, remember that $\ntispace^\sigma(n^t,n^s)$ denotes the
class of problems over $\sigma$-structures $\langle [n],\sigma
\rangle$ recognizable by an \nram\ that uses computation register
contents $O(n^s)$ and works in time $O(n^t)$ and space $O(n^s)$.

\begin{definition} \label{codeS}
For  any unary $\left\{ f \right\}$-structure $S=\langle [m],f
\rangle ,f:[m]\rightarrow[m]$, let $code(S)=\langle [n],g \rangle$
denote the structure of signature $\sigma_d = \left\{
g_0,\ldots,g_{d-1} \right\}$, where every $g_i$ is of arity $d$,
defined by $(1-2)$:
\begin{enumerate}
  \item $ n-1 = \lceil m^{1/d} \rceil$, i.e. $(n-2)^d < m \leq (n-1)^d$;
  \item $g:[n]^d \rightarrow [n]^d$ is such that $g=(g_0,\ldots,g_{d-1})$, where $g_i: [n]^d \rightarrow [n]$, and
   \begin{enumerate}
     \item[2.1.] if $f(a)=b$ for any $a,b<m$, then $g_i(a_0,a_1,\dots,a_{d-1})=b_i$
     where $a_i,b_i$ are the respective $i^{th}$ digits of $a,b$ in base $n-1$,
     that means $a= \sum_{i<d} a_i (n-1)^i$ and $b= \sum_{i<d} b_i (n-1)^i$ with $a_i,b_i<n-1$, and
     \item[2.2.] $g_i(a_0,a_1,\dots,a_{d-1})=n-1$ if $(a_0,a_1,\dots,a_{d-1}) \in [n]^d$ is
     not the list of $(n-1)$-digits of any integer smaller than $m$.
   \end{enumerate}
\end{enumerate}
Let ${\cal P}^{code}=\{ code(S) : S \in {\cal P} \}$.
\end{definition}

The following remarks are essential.

\begin{remark}
$S$ and $code(S)$ have about the same size, i.e. $size(code(S))=\Theta (n^d)=\Theta (m)=\Theta (size(S))$.
\end{remark}

\begin{remark} \label{corresp s code s}
The  correspondence $S \mapsto code(S)$ is one-one and is easily
computable as its converse is because if $S'=\langle [n],g \rangle
= code(S)$ for $S=\langle [m],f \rangle$ then we have
\begin{equation} \label{ScodeS}
m=\sharp \left\{ (a_0,a_1,\dots,a_{d-1}) \in [n-1]^d : g_0(a_0,a_1,\dots,a_{d-1})<n-1 \right\}
\end{equation}
\end{remark}

Here is our technical lemma:

\begin{lemma} \label{lemma}
For  any fixed numbers $t,s$ such that $t \geq  s \geq 1$ and $t
\geq d \geq 1$, ${\cal P}\in \ntispace(m^{t/d},m^{s/d})$ implies
${\cal P}^{code} \in \ntispace^{\sigma_d}(n^t,n^s)$.
\end{lemma}

\begin{proof}
Under the hypothesis, let us give an\\
\begin{description}
 \item[] {\bf Algorithm for recognizing the problem ${\mathbf {\cal P}^{code}}$}
 \begin{description}
  \item[] {\bf Input : }a $d$-ary structure $S'=\langle [n],g \rangle$.
  \item[] {\bf begin}
  \begin{itemize}
    \item compute $m$ with the expression (\ref{ScodeS}) of Remark \ref{corresp s code s};
    \item check that all the conditions ($1$-$2$) of Definition \ref{codeS} are satisfied,
    that means $S'=code(S)$ for some (unique) unary structure $S=\langle [m],f \rangle$
    with $n-1 = \lceil m^{1/d} \rceil$;
    \item on the input $S$, simulate the running of the \nram\ that recognizes \pp\ in
    time $O(m^{t/d})=O(n^t)$ and space $O(m^{s/d})=O(n^s)$. (Note that this \nram\ only
    uses register contents $O(m)=O(n^d)$.)
  \end{itemize}
  \item[] {\bf end}\\
 \end{description}
\end{description}

This proves ${\cal P}^{code} \in \ntispace^{\sigma_d}(n^t,n^s)$.
\qed
\end{proof}

\subsection{Completeness result}

Let us now present our logical problems, denoted $\lc(t,s)$, where $t,s$ are fixed integers, $t \geq s \geq 1$.

\begin{definition}
An  {\em $[n]$-formula $F$ of signature $\sigma$} is a
quantifier-free first-order $\sigma$-formula where the variables
are replaced by integers in $[n]$. Let {\em $length(F)$} denote
the number of occurrences of integers, $\sigma$-symbols,
equalities, and connectives in $F$.
\end{definition}

\noindent
{\bf Example: }$g(h(1,0),2)=h(3,1)$ is an atomic $[4]$-formula of signature $\{ g,h \}$ and of length~$9$.

\begin{description}
  \item[{\bf Problem} {\sc Layered-Constraints$(t,s)$}]
  \begin{description}
    \item
    \item[{\bf Instance :}]
       \begin{itemize}
         \item an integer $n$;
         \item a non-empty set $\cal OP$ of $t$-ary operators $[n]^t \rightarrow [n]$,
         each explicitly given by its table;
         \item a conjunction of $[n]$-formulas $F = F_1 \wedge \ldots \wedge F_l$,
         each $F_i$ of size at most $n^s$ and of signature $\nu_i$,
         where $\nu_i \cap \nu_j \subseteq {\cal OP}$ if $|i-j|>1$ ($\ast$).
       \end{itemize}
    \item[{\bf Question :}] Is $F$ satisfiable?
  \end{description}
\end{description}

\begin{convention}
To satisfy automatically condition $(\ast)$, it is natural to
partition the overall signature of $F$ as  $\tau_0 \dot \cup
\tau_1 \dot \cup \ldots \dot \cup \tau_l \dot \cup {\cal OP}$,
where $\dot \cup$ denotes the disjoint union, with
$\nu_i=\tau_{i-1} \cup \tau_i$,  $\tau_0=\emptyset$. The $j^{th}$
symbol of $\tau_i$ is denoted $f_j^i$.
\end{convention}

\begin{remark}
The size of the input is $m \geq n^t$;
\end{remark}

\begin{proposition} \label{appartenance}
{\sc Layered-Constraints$(t,s)$} is in $\ntispace(m,m^{s/t})$.
\end{proposition}

\begin{proof}
Let $(n,{\cal OP},F_1 \wedge \ldots \wedge F_l)$ be an instance of
$\lc(t,s)$. First, we consider formula $F_1$ and prove that it is
coherent with $\cal OP$. Remember that we use a nondeterministic
\ram. Each part of the formula is checked in the following way.
\begin{itemize}

  \item If it is of the form $F'_i \vee F'_j$, with $F'_i$ and $F'_j$
  subformulas of $F_1$, then we nondeterministically choose $F'_i$
  {\em or} $F'_j$ and check its coherence.

  \item If it is of the form $F'_i \wedge F'_j$, with $F'_i$ and $F'_j$
  subformulas of $F_1$, then we check if $F'_i$ {\em and} $F'_j$ are coherent.

  \item If it is of the form $f(\alpha)=\beta$, with $f$ a symbol
  of $\nu_1$, $\beta \in [n]$ and $\alpha \in [n]^k$ where $k$ is
  the arity of $f$, then we set $f(\alpha)=\beta$ and store it in the memory.

  \item If it is of the form $f(\alpha)=g(\beta)$, with $f,g$
  symbols of $\nu_1$ and $\alpha \in [n]^k,\beta \in [n]^m $
  where $k$ is the arity of $f$ and $m$ the arity of $g$, then we
  nondeterministically choose an interpretation for $f$ in $\alpha$
  and give the same value to $g$ in $\beta$.

  \item The same applies if we have compositions of the
  form $f \left( g_1(\ldots), \ldots, g_k(\ldots) \right)$,
  whatever the arity of the functions involved.

  \item Whenever an element of $\cal OP$ appears, we take the interpretation given in the input.
\end{itemize}

Note that, as the size of $F_1$ is at most $n^s$, we don't need
more than $n^s$ registers to store the values we need. (Note also
that a symbol is identified with the signature in which it
appears, so it is easy to see if a formula uses a forbidden
symbol.) Once this is done for formula $F_1$, we sort all the
values we have given to the functions, there are at most $|F_1|$
so it can be done in time and space $O(|F_1|)$ (see
\cite{grandjean96}), and we check that if there are twice (or
more) the same symbol in the same value, then the same
interpretation is given every time. All this is done in time
$O(|F_1|)$ and space $O(n^s)$. We do the same thing for $F_2$, but
as there may be common symbols in both formulas $F_1$ and $F_2$,
we check that they were given the same value (as both lists are
sorted, it can be done at the same time as the check for repeated
occurrences). This again is done in time $O(|F_1|+|F_2|)$ and
space $O(n^s)$. Now remember that symbols in $F_1$ no longer occur
in the other formulas $F_i$ for $i>2$, so we can forget their
interpretations as they can no longer bring incoherence. The
memory thus freed will be used to store the values that appear in
$F_3$, and so on until we check the coherence of $F_l$.

Overall, the same memory space $O(n^s)=O(m^{s/t})$ is always
recycled and the time needed is $O(\sum_{1 \leq i \leq l}
|F_i|)=O(m)$, where $m$ is the size of the input. So $\lc(t,s) \in
\ntispace(m,m^{s/t})$.  \qed
\end{proof}

The following theorem essentially expresses the completeness of the problem \\
$\lc(t,s)$ in the class \ntispace$(m,m^{s/t})$.

\begin{theorem} \label{big th}
For all integers $t,s$, $t \geq s \geq 1$, and all functions $T,S$:
\begin{enumerate}
  \item[(i)] $\lc(t,s) \in \dtisp(T(O(m)),S(O(m)))$ if and only if \\
$\ntispace(m,m^{s/t}) \subseteq \dtisp(T(O(m)),S(O(m)))$;
  \item[(ii)] $\lc(t,s) \in \contisp(T(O(m)),S(O(m)))$ if and only if \\
$\ntispace(m,m^{s/t}) \subseteq \contisp(T(O(m)),S(O(m)))$.
\end{enumerate}
\end{theorem}

\begin{proof}[Sketch of]
Proposition \ref{appartenance}  yields the {\em if} implication.
We will prove  the {\em only if} part of this theorem for the case
$s=1,t=2$, the general case being an easy generalization of this
particular one.

Let ${\cal P}$ be a problem  in $\ntispace(m,m^{1/2})$. Then by
Lemma \ref{lemma}, ${\cal P}^{code} \in
\ntispace^{\sigma_2}(n^2,n)$, with $\sigma_2= \left\{ g_0,g_1
\right\}$, and by Theorem \ref{th main} there exists a formula
$\varphi$ in $\eso^{\sigma_2}(1,1)$ such that for each integer
$n>0$ and each $\sigma_2$-structure $\langle [n], g_0,g_1 \rangle$
we have
\begin{equation} \label{final proof 1}
\langle [n], g_0,g_1 \rangle \in {\cal P}^{code} {\mbox { iff }} \langle [n],g_0,g_1,succ,0 \rangle \models \varphi
\end{equation}

Let $\varphi \equiv \exists {\bar f}\  \forall x\  \forall y\
\psi(x,y)$ where $y$ is the iteration variable and $\psi$ is
quantifier-free. Without loss of generality, assume that ${\bar
f}$ consists of function symbols $f_i$ of arity $2$. The idea
consists in unfolding the first-order part of $\varphi$ on the
domain $[n]$. This gives the equivalent $[n]$-formula
$$
\bigwedge_{b<n} \left[ \bigwedge_{a<n} \psi (a,b) \right]
$$
which is also equivalent to  the conjunction $\bigwedge_{b<n} F_b$
where $F_b=\left[ \bigwedge_{a<n} \psi_b (a) \right]$ and
$\psi_b(a)$ denotes the formula $\psi(a,b)$ in which each term
$f_i(a,b)$ (resp. $f_i(a,succ(b))$), for $f_i \in \bar f$, is
replaced by $f_i^b(a)$ (resp. $f_i^{b+1}(a)$). In other words,
each \eso\ function symbol $f_i \in \bar f$ (of arity $2$) is
replaced by $n$ function symbols $f_i^b$ ($b<n$) of arity $1$. By
construction, we have $\langle  [n],g_0,g_1,succ,0 \rangle \models
\varphi$ iff the $[n]$-formula $F \equiv  \bigwedge_{b<n} F_b$ is
coherent with the tables of functions $g_0$ and $g_1$. Note that,
by construction, if $|b - b'|>1$, we have $signature(F_b) \cap
signature(F_{b'}) \subseteq \{g_0,g_1\}$.

There remains a technical problem. $(n,\sigma_2,F)$ is not exactly
an instance of the problem $\lc(2,1)$ since $length(F_b)=kn=N$ for
each $b<n$, where $k=length(\psi)$. Therefore, we ``linearly pad''
our instance into an ``equivalent'' instance over $[N]$ by
completing the tables of $g_0$ and $g_1$ on the domain $[N]$,
under the names $g_0^{(N)}$ and $g_1^{(N)}$ respectively: we add
the values $g_i^{(N)}(a_0,a_1)=(0,0)$ whenever  $a_0$ or $a_1$
belongs to $[N] \setminus [n]$. We obtain an instance $(N,{\cal
OP},F)$ of $\lc(2,1)$ with $N=kn$ and ${\cal OP}= \left\{
g_0^{(N)},g_1^{(N)} \right\}$ such that (by (\ref{final proof 1}))
$$
\langle [n],g_0,g_1 \rangle \in {\cal P}^{code} {\mbox { iff }}
(N,{\cal OP},F) \in \lc(2,1)
$$
Let us recapitulate the properties of our reduction of any problem
$\cal P$ of $\ntispace(m,m^{1/2})$ to $\lc(2,1)$.
\begin{itemize}

  \item{\bf It is correct:}\\
For any $m$ and any input unary structure $S=\langle [m],f
\rangle$, if $S'=code(S)=\langle [n],g_0,g_1 \rangle$:
$$
S \in {\cal P} {\mbox { iff }} S' \in {\cal P}^{code} {\mbox { iff
}} (N,{\cal OP},F) \in \lc(2,1)
$$

  \item{\bf It is linear-sized:}
$$
length(F) =
\Theta(N^2)=\Theta(n^2)=\Theta(size(S'))=\Theta(size(S))
$$
and similarly for $size({\cal OP})$.

  \item{\bf It yields the ``only if'' implication of the Theorem \ref{big th}(i):}\\
Let $\cal R$ be a (deterministic) \ram\ that decides \\
$\lc(2,1)$ in time $O(T(O(m)))$ and space $O(S(O(m)))$. The \ram\
$\cal R'$ with the following program decides the problem $\cal P$.
\begin{description}
  \item[] {\bf Input : }a unary $\left\{ f \right\}$-structure $S=\langle [m],f \rangle$.
  \item[] {\bf begin}
  \begin{itemize}
    \item Compute $n=\lceil m^{1/2} \rceil$ and $N=kn$.
    \item Simulate running $\cal R$ on input $(N,{\cal OP},F)$ without
storing it nor $S'=code(S)$. Whenever $\cal R$ needs to read an
operator value $g_i^{(N)}(a_0,a_1)=b_i$ ($i=0$ or $i=1$), such
that, e.g., $a_0<n-1,a_1<n-1$ and $a_0+a_1(n-1)<m$, $\cal R'$
reads the value $v=f(a_0+a_1(n-1))$ in its input $S$ and compute
$b_0=v \bmod (n-1)$ or $b_1 = \lfloor v/(n-1) \rfloor$. Whenever
$\cal R$ needs to read a symbol in $F$, the easy structure of $F
\equiv F_0 \wedge F_1 \wedge \dots \wedge F_{n-1}$ allows $\cal
R'$ to compute that symbol in constant time and constant space.
  \end{itemize}
  \item[] {\bf end}
 \end{description}
Since the reduction $S \mapsto (N,{\cal OP}, F)$ is linear-sized,
$\cal R'$ decides $\cal P$ within time $O(T(O(m)))$ and space
$O(S(O(m)))$ as required. The proof of part (ii) of Theorem
\ref{big th} is similar. \qed

\end{itemize}

\end{proof}

\section{Corollaries: time-space lower bounds} \label{lower bounds}

The completeness results for  the problems $\lc(t,s)$ obtained in
the previous section yield some lower bounds for those problems
because of several separation results proved by Fortnow et al.
\cite{fortnowmelkebeek00} that we reformulate as follows.

\begin{theorem}[Fortnow-Melkebeek 2000: See Corollary 4.8, Corollary 3.23, Corollary 3.22, respectively]
\begin{itemize}
  \item $\ntispace(m,m^{0.619}) \not \subseteq \dtisp(m^{1.618},m^{o(1)})$
  \item $\ntispace(m,m^{3/4}) \not \subseteq \contisp(m^{1.4},m^{o(1)})$
  \item $\ntispace(m,m^{4/5}) \not \subseteq \contisp(m^{5/4},m^{1/10})$
\end{itemize}
\end{theorem}

\begin{corollary}
\begin{itemize}
  \item $\lc(8,5) \in \ntispace(m,m^{5/8}) \setminus \dtisp(m^{1.618},m^{o(1)})$
  \item $\lc(3,2) \in \ntispace(m,m^{2/3}) \setminus \dtisp(m^{1.618},m^{o(1)})$
  \item $\lc(4,3) \in \ntispace(m,m^{3/4}) \setminus \contisp(m^{1.4},m^{o(1)})$
  \item $\lc(5,4) \in \ntispace(m,m^{4/5}) \setminus \contisp(m^{5/4},m^{1/10})$
\end{itemize}
\end{corollary}

\section{Conclusion and open problems}

It is well-known that most  natural \np-complete problems are in
\nlin\ and that many of them, e.g. \sat, are complete in
nondeterministic quasi-linear time. Moreover, in this paper we
have shown that significant \np-complete problems, mainly many
problems over planar graphs and some problems over numbers, belong
to $\ntispace(n,n^q)$, for some $q<1$, and specifically to
$\ntispace(n,n^{1/2})$. This improves the known upper bound
$\dtime(2^{O(n^{1/2})})$ for those problems. Thanks to a logical
description of nondeterministic {\em polynomial} time-space
classes we have exhibited, for any integers $s,t$, $t \geq s \geq
1$, a problem, denoted $\lc(t,s)$, that is complete in
$\ntispace(n,n^{s/t})$ via linear reductions. This is a very
precise and nontrivial result. The main open challenge would be to
discover ``more natural'' complete problems in such classes,
mainly in $\ntispace(n,n^{1/2})$, even via quasi-linear
reductions. Further, in order to prove some complexity lower
bound, such a result should be completed by some separation result
for $\ntispace(n,n^{1/2})$ that would be similar to those obtained
by \cite{fortnowmelkebeek00} for any class $\ntispace(n,n^q)$,
with $q$ greater than $0.619$ (the golden ratio).

\bibliographystyle{plain}
\bibliography{biblio}

\end{document}